\documentclass[10pt,aps,pra,twocolumn,showpacs,superscriptaddress]{revtex4-1}

\usepackage{graphicx}
\usepackage{amsmath,amssymb,amsfonts,dsfont}
\usepackage{color}
\usepackage[normalem]{ulem} 
\usepackage{multirow}
\usepackage{array}
\usepackage{upgreek}
\usepackage{mathrsfs}
\usepackage{dblfloatfix}
\usepackage[table]{xcolor}
\usepackage{float}
\restylefloat{table}

\newcommand{\bra}[1]{\langle{#1}|}
\newcommand{\ket}[1]{|{#1}\rangle}

\begin{document}

\title{Engineering Framework for Optimizing Superconducting Qubit Designs}

\author{\mbox{Fei Yan}}
\thanks{yanf2020@mail.sustech.edu.cn}
\altaffiliation{Current address: Shenzhen Institute for Quantum Science and Engineering, Southern University of Science and Technology, Shenzhen, Guangdong 518055, China}
\affiliation{Research Laboratory of Electronics, Massachusetts Institute of Technology, Cambridge, MA 02139, USA}
\author{Youngkyu Sung}
\affiliation{Research Laboratory of Electronics, Massachusetts Institute of Technology, Cambridge, MA 02139, USA}
\affiliation{Department of Electrical Engineering and Computer Science, Massachusetts Institute of Technology, Cambridge, MA 02139, USA}
\author{Philip Krantz}
\affiliation{Research Laboratory of Electronics, Massachusetts Institute of Technology, Cambridge, MA 02139, USA}
\author{Archana Kamal}
\altaffiliation{Current address: Department of Physics and Applied Physics, University of Massachusetts Lowell, Lowell, MA 01854, USA}
\affiliation{Research Laboratory of Electronics, Massachusetts Institute of Technology, Cambridge, MA 02139, USA}
\author{David K. Kim}
\affiliation{MIT Lincoln Laboratory, 244 Wood Street, Lexington, MA 02421, USA}
\author{Jonilyn L. Yoder}
\affiliation{MIT Lincoln Laboratory, 244 Wood Street, Lexington, MA 02421, USA}
\author{Terry P. Orlando}
\affiliation{Research Laboratory of Electronics, Massachusetts Institute of Technology, Cambridge, MA 02139, USA}
\author{Simon Gustavsson}
\affiliation{Research Laboratory of Electronics, Massachusetts Institute of Technology, Cambridge, MA 02139, USA}
\author{\mbox{William D. Oliver}}
\affiliation{Research Laboratory of Electronics, Massachusetts Institute of Technology, Cambridge, MA 02139, USA}
\affiliation{Department of Electrical Engineering and Computer Science, Massachusetts Institute of Technology, Cambridge, MA 02139, USA}
\affiliation{MIT Lincoln Laboratory, 244 Wood Street, Lexington, MA 02421, USA}
\affiliation{Department of Physics, Massachusetts Institute of Technology, Cambridge, MA 02139, USA}

\begin{abstract}

	Superconducting quantum technologies require qubit systems whose properties meet several often conflicting requirements, such as long coherence times and high anharmonicity.
	%
	%
	Here, we provide an engineering framework based on a generalized superconducting qubit model in the flux regime, which abstracts multiple
    circuit design parameters and thereby supports design optimization across multiple qubit properties.
	%
	We experimentally investigate a special parameter regime which has
	both high anharmonicity ($\sim\!1$\,GHz) and
	long quantum coherence times ($T_1\!=\!40\!-\!80\,\mathrm{\mu s}$ and $T_\mathrm{2Echo}\!=\!2T_1$).
	%
\end{abstract}

\maketitle


Since the first direct observation of quantum coherence in a superconducting qubit more than 20 years ago \cite{Nakamura-Nature-1999}, many variants have been designed and studied \cite{krantz2019quantum},
 such as the Cooper-pair box (CPB) \cite{Nakamura-Nature-1999}, the persistent-current flux qubit (PCFQ) \cite{Orlando-PRB-1999,Mooij-Science-1999}, the transmon \cite{Koch-PRA-2007,Paik-PRL-2011}, the fluxonium \cite{Manucharyan-Science-2009}, and the capacitively-shunted flux qubit (CSFQ) \cite{Steffen-PRL-2010,Yan-NComms-2016}.
%
%
These superconducting qubit designs were usually categorized according to the ratio between the effective charging energy $E_\mathrm{C}$ and Josephson energy $E_\mathrm{J}$, into the charge ($E_\mathrm{J} \leq E_\mathrm{C}$) or flux ($E_\mathrm{J} \gg E_\mathrm{C}$) regime \cite{Clarke-Nature-2008}.

\begin{figure}
	\begin{center}
		\vspace{-0.0cm}
		\includegraphics[scale=1]{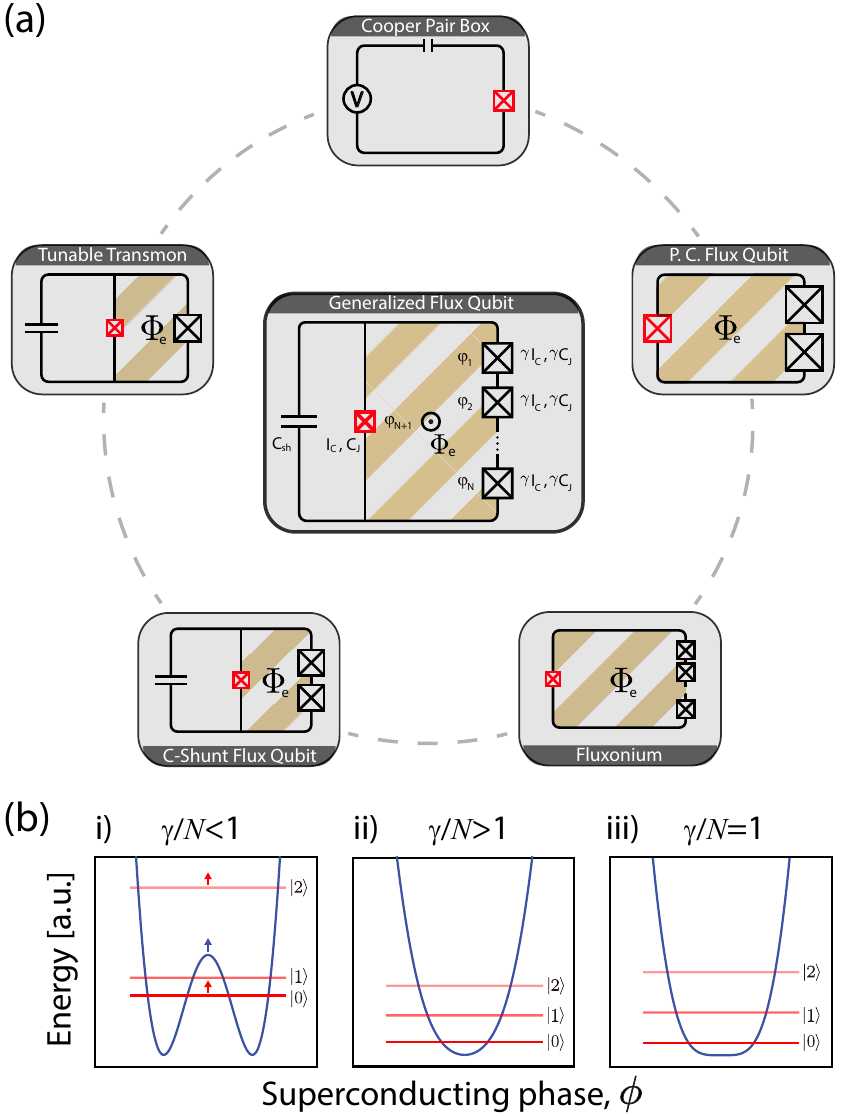}
		\caption[]
		{
			(\textbf{a})
			In the center is the circuit diagram of the generalized flux qubit, where the principal junction (red box), parameterized by its critical current $I_\mathrm{c}$ and junction capacitance $C_\mathrm{J}$, is shunted by both a capacitor $C_\mathrm{sh}$ and an array of $N$ larger junctions ($\gamma$ times in size).
			%
			The surrounding circuits are demonstrated qubit variants, including the Cooper pair box and qubits in the flux regime: tunable transmon, persistent-current flux qubit, capacitively-shunted flux qubit, and fluxonium.
			From geometry, they can be viewed as reduced versions of the generalized flux qubit circuit.
			Note that the tunable transmon has different topology in its Hamiltonian. The smaller junction (red) should not be regarded as the principle junction as in the other cases.
			(\textbf{b})
			Exemplary potential profile and lowest energy levels in the \textbf{\romannumeral 1)} fluxon, \textbf{\romannumeral 2)} plasmon, and \textbf{\romannumeral 3)} quarton regime.
			In the left plot, red arrows indicate how $\ket{0}$ and $\ket{2}$ moves with increasing center barrier (blue arrows), or decreasing $\gamma/N$.
		}
		\label{fig:circuit_main}
	\end{center}
\end{figure}

The CPB (Fig.~\ref{fig:circuit_main}), a representative in the charge regime, provides large anharmonicity that facilitates fast gate operations.
However, strong background charge noise limits its coherence time \cite{Nakamura-Nature-1999}, and the dispersion from quasiparticle tunneling causes severe frequency instability \cite{schreier2008suppressing}.
Likewise, qubits in the flux regime, including the transmon, PCFQ, CSFQ, fluxonium and the  rf-SQUID qubit, have been studied extensively as potential elements for gate-based quantum computing \cite{Kelly-Nature-2015,Ofek-Nature-2016,corcoles2015demonstration,Riste-NComms-2015}, quantum annealing \cite{Johnson-Nature-2011,Barends-Nature-2016}, quantum simulations \cite{Barends-NComms-2015} and many other applications, largely due to the flexibility in engineering their Hamiltonians and due to 
their relative insensitivity to charge noise.
%
%
%
%

In this work, we provide an engineering framework based on a generalized flux qubit (GFQ) model which accommodates most (if not all) contemporary qubit variants.
The framework facilitates an understanding of how key qubit properties are related to circuit parameters.
The increased complexity, i.e., the use of both a shunt capacitor and an array of Josephson junctions, enables better control over coherence, anharmonicity and qubit frequency.
%
As an example of implementing this framework, we experimentally demonstrate a special parameter regime, the ``quarton'' regime, named after its quartic potential profile.
In comparison with other state-of-the-art designs, the quarton can simultaneously maintain a desirable qubit frequency ($3\!-\!4$\,GHz), large anharmonicity ($\sim\!1$\,GHz), and high coherence ($T_1\!=\!40\!-\!80\,\mathrm{\mu s}$, $T_\mathrm{2Echo}\!=\!2T_1$).
Such a configurable energy level structure 
is advantageous with respect to the problem of frequency crowding in highly connected qubit systems.
%
We show experimentally that quarton qubits with as few as 8 and 16 array junctions and with a much smaller shunt capacitor allow for a compact design, promising better scalability and reproducibility.
%

In the GFQ circuit with $N$ array junctions (Fig.~\ref{fig:circuit_main}), each junction is associated with a gauge-invariant (branch) phase $\varphi_i$.
These phases satisfy the fluxoid quantization condition, $\sum_{k=1}^{N+1}\!\varphi_k \,+\, \varphi_\mathrm{e} = 2\pi z$ ($z\in\mathbb{Z}$), where $\varphi_\mathrm{e} = 2\pi \Phi_\mathrm{e}/\Phi_0$. $\Phi_\mathrm{e}$ is the external magnetic flux threading the qubit loop and $\Phi_0=h/2\mathrm{e}$ is the superconducting flux quantum.
%
%
Although the full Hamiltonian is $N$-dimensional,  the symmetry among the array junctions allows the full Hamiltonian  to  be approximated by a one-dimensional Hamiltonian; and this dimension coincides with a lower-energy mode that describes the qubit \cite{Ferguson-PRX-2013}.
This  one-dimensional  Hamiltonian is given by
%
\begin{align} \label{eq:H_plus}
\mathcal{H} &= - 4 E_\mathrm{C} \, {\partial_{\phi}}^2 + E_\mathrm{J} \Big( - \gamma N \cos(\phi/N) - \cos(\phi + \varphi_\mathrm{e}) \Big)  \;,
\end{align}
where $\phi=\varphi_1+\varphi_2+...+\varphi_N$ is the phase variable of the qubit mode, and $\gamma$ is the size ratio between the array junction and the smaller principle junction.
The effective charging energy is $E_\mathrm{C} = \mathrm{e}^2/2C_\Sigma$, where $C_\Sigma = C_\mathrm{sh} + C_\mathrm{J} + \gamma C_\mathrm{J}/N + C_\mathrm{g}$ is the total capacitance across the principal junction.
$C_\mathrm{g}$ is the correction from stray capacitances from superconducting islands to ground, which may become significant for large $N$ \cite{Ferguson-PRX-2013,Viola-PRB-2015}.
The principal junction has a Josephson energy $E_\mathrm{J} = I_\mathrm{c} \Phi_0/2\pi$.
In the $E_\mathrm{J}\!\gg\!E_\mathrm{C}$ limit, $\phi$ is well-defined and has small quantum fluctuations.

Such a multi-junction qubit (total junction number $\geqslant3$) achieves the best coherence when biased at $\varphi_\mathrm{e}=\pi$, where the qubit frequency is (at least) first-order insensitive to flux fluctuations.
At this working point, we may expand the potential part of Eq.~(\ref{eq:H_plus}) to fourth order,
\begin{align} \label{eq:H_plus_4th}
\mathcal{H} &= - 4 E_\mathrm{C} \, {\partial_{\phi}}^2 + E_\mathrm{J} \Big( \frac{\gamma/N-1}{2} \phi^2 + \frac{1}{24} \phi^4 \Big) \;,
\end{align}
where we have assumed $N^3\!\gg\!\gamma$.
Depending on the value of $\gamma/N$, the problem can be categorized into one of the three regimes illustrated in Fig.~\ref{fig:circuit_main}(b).
The fluxon regime (\textbf{\romannumeral 1}), $1\!<\!\gamma\!<\!N$, was first demonstrated in the traditional PCFQ with $N=2$ \cite{VanDerWal-Science-2000}, where the potential assumes a double-well profile, providing strong anharmonicity.
The fluxonium extends the case to $N\!\approx\!100$ \cite{Pop-Nature-2014}.
%
The energy eigenstates can be treated as hybridized states via quantum-mechanical tunneling between neighboring wells.
The plasmon regime (\textbf{\romannumeral 2}), $\gamma\!>\!N$, was explored in the CSFQ, where the potential assumes a single-well profile. A leading quadratic term and a minor quartic term lead to weak anharmonicity, though the CSFQ is still more anharmonic than the transmon due to partial cancellation of the quadratic term \cite{Yan-NComms-2016}.
The quarton regime (\textbf{\romannumeral 3}), $\gamma \approx N$, approximates the problem to a particle in a quartic potential.
%
%
As we will show later, the quarton design has desirable features in its energy level configuration.
We notice that a similar design with $\gamma\approx N=3$ was used in a parametric amplifier, but it operates at non-degenerate bias to exploit the cubic potential term and, to the contrary, eliminate the quartic one \cite{frattini20173}.

\begin{figure}
	\begin{center}
		\vspace{-0.0cm}
		\hspace{0cm}
		\includegraphics[scale=1]{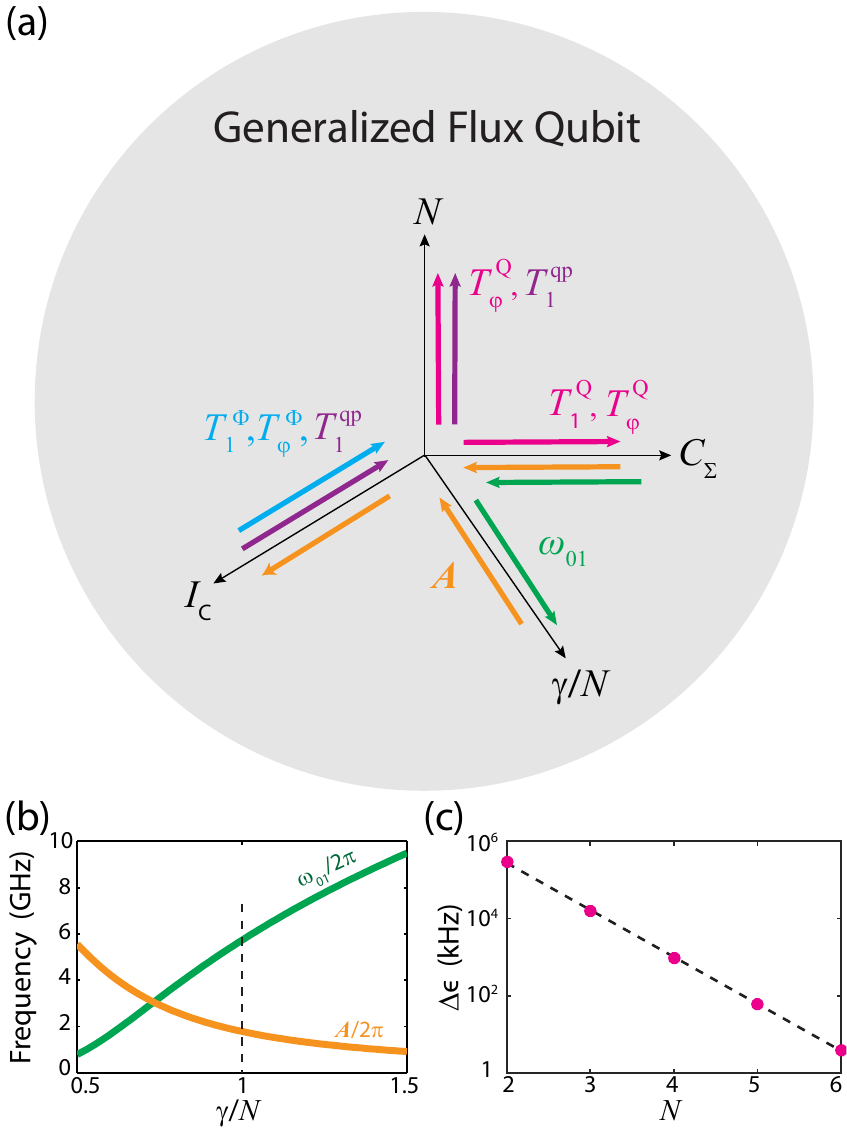}
		\caption[]
		{
			\textbf{(a)} Illustration of how relevant qubit properties depend on key control variables in the GFQ framework.
			For example, the arrows along the $\gamma/N$ axis indicate that anharmonicity $\mathcal{A}$ is a decreasing function of $\gamma/N$, while qubit frequency $\omega_{01}$ is increasing.
			Other included qubit properties are energy relaxation ($T_1$) and pure dephasing ($T_\varphi$) due to flux noise ($\Phi$), charge noise (Q) or quasiparticle tunneling (qp).
			%
			%
			\textbf{(b)} The $\gamma/N$-dependence of the qubit frequency and anharmonicity, simulated from a 1D model with $I_\mathrm{c}=40$\,nA, $C_\mathrm{sh}=20$\,fF, $C_\mathrm{J}=1$\,fF. Dashed line indicates the quarton case.
			\textbf{(c)} The $N$-dependence of the size of the charge dispersion in the quarton case, simulated from the full $N$-dimensional model with the same parameters as in \textbf{(b)}.
			The dashed line depicts an empirical dependence: $\sqrt{16 E_\mathrm{J} E_\mathrm{C}} \exp\left({-\sqrt{N^{2.5} E_\mathrm{J} / 7 E_\mathrm{C}}}\right)$.
		}
		\label{fig:parameter_tradespace}
	\end{center}
\end{figure}

%

To find qubit designs with predetermined desirable properties, we may expand the parameter space beyond $E_\mathrm{J}/E_\mathrm{C}$, to include
$I_\mathrm{c}$, $C_\Sigma$, $N$ and $\gamma/N$ as the four independent design parameters.
Our engineering framework provides an abstraction that captures the underlying physics to develop a set of rules or guidelines by which one can understand the parameter-property tradespace, as illustrated in Fig.~\ref{fig:parameter_tradespace}(a).
%
%
%
%
%
%
%
%
%
%
Two of the circuit parameters, $I_\mathrm{c}$ and $C_\Sigma$, have been studied extensively.
In general, a lower $I_\mathrm{c}$ and a higher $C_\Sigma$ are preferred for reducing sensitivity to flux and charge noise respectively.
The energy level structure is also generally sensitive to their values, depending on the specific case.
%
In the following, we focus on the discussion of the other two quantities, $\gamma/N$ and $N$.
%

First, we consider $\gamma/N$ as an independent variable instead of $\gamma$, because the Hamiltonian in Eq.~(\ref{eq:H_plus_4th}) is parameterized by $E_\mathrm{J}$, $E_\mathrm{C}$, and $\gamma/N$.
We find that a smaller $\gamma/N$ leads to a smaller qubit frequency and a larger anharmonicity, except for certain cases such as very small $N$ or $\gamma\approx1$.
An example is shown in Fig.~\ref{fig:parameter_tradespace}(b).
An intuitive explanation is as follows.
With a symmetric potential profile, the wavefunctions of the ground state $\ket{0}$ and the second-excited state $\ket{2}$ have even parity while the excited state $\ket{1}$ has odd parity.
Reducing $\gamma/N$ will raise the potential energy around $\phi = 0$, pushing up $\ket{0}$ and $\ket{2}$ due to their non-zero amplitudes at $\phi = 0$.
In contrast, the odd-parity $\ket{1}$ state is unaffected, leading to a smaller $\omega_{01}$ and a greater $\omega_{12}$.

At the critical value $\gamma/N=1$, the quadratic term in Eq.~(\ref{eq:H_plus_4th}) is canceled, resulting in a quartic potential.
We can find the solutions numerically with  $E_n = \lambda_n ( \frac{2}{3} E_\mathrm{J} {E_\mathrm{C}}^2 )^{1/3}$.
For the lowest three levels, we find $\lambda_0=1.0604$, $\lambda_1=3.7997$ and $\lambda_2=7.4557$.
Note that $\lambda_2-\lambda_1\approx \frac{4}{3}(\lambda_1-\lambda_0)$, suggesting that the anharmonicity of the quarton qubit is about 1/3 of its qubit frequency.
%
%
%
This interesting finding is useful in practice, as it is common to operate qubits in the frequency range of 3-6\,GHz, in part for better qubit initialization ($\omega_{01}\!\gg\!k_\mathrm{B}T/\hbar$), and in part for compatibility with high-performance microwave control electronics, although, exceptions exist using non-adiabatic control \cite{campbell2020universal,zhang2020universal}.
One third of the qubit frequency gives 1-2\,GHz anharmonicity, sufficient for suppressing leakage to non-computational states and alleviating the frequency-crowding problem, so that higher single- and two-qubit gate fidelities are achievable.

Second, we find that the charge dispersion that causes qubit frequency instability and dephasing can be efficiently suppressed by increasing $N$.
The size of the charge dispersion $\Delta\epsilon$ can be estimated from the tight-binding hopping amplitude between neighboring lattice sites in the potential landscape, which becomes exponentially small with respect to the height of the inter-lattice barrier.
In the example of a single junction, it was shown that $\Delta\epsilon \propto \exp(-\sqrt{8 E_\mathrm{J}/E_\mathrm{C}})$ \cite{Koch-PRA-2007}.
It is more complicated for flux qubits due to the multi-dimensionality of the Hamiltonian.
In the quarton, a simulation of the full $N$-dimensional Hamiltonian shows that the suppression is even more efficient, because the barrier height scales with $N^2$ [Fig.~\ref{fig:parameter_tradespace}(c)].
Typically, $N\geqslant6$ is sufficient to suppress charge dispersion down to the kilohertz level.
This implies that it is not necessary to use $\sim\!\!100$ junctions to suppress the charge noise for better coherence~\cite{Manucharyan-Science-2009,Pop-Nature-2014}.
%
A more compact and easier-to-fabricate design can be done with the quarton.

There are other advantages of using a junction array.
For example, $T_1$ relaxation due to quasiparticle tunneling across an array junction may be improved by increasing $N$, since the corresponding matrix element $\bra{0} \sin (\frac{\varphi_i}{2}) \ket{1} \approx \bra{0} \sin (\frac{\phi}{2N}) \ket{1}$ scales as $1/N$ (it vanishes at $\varphi_\mathrm{e}=\pi$ for tunneling across the principal junction) \cite{Catelani-PRB-2011}.
However, it is also important to limit the array size, since more junctions lead to more decohering channels from the parasitic capacitance to ground and the Aharonov-Casher effect, as well as keeping the qubit loop area small to avoid excess flux noise.
In general, a trade-off has to be found.
Such an example is discussed within the context of optimizing coherence time given only a few noise sources \cite{mizel2019rightsizing}.

The enhanced controllability over qubit properties in the GFQ model allows more flexibility in qubit design.
For example, the transmon requires a large shunt capacitance to suppress charge dispersion.
This unavoidably lowers $E_\mathrm{C}$ and anharmonicity, as well as increases the qubit footprint.
With the introduction of the junction array in the GFQ, one gains more freedom in configuring the qubit frequency and anharmonicity while, at the same time, the junction array helps suppress charge dispersion.
%

\begin{figure}[ht!]
	\begin{center}
		\vspace{-0.0cm}
		\includegraphics[scale=1]{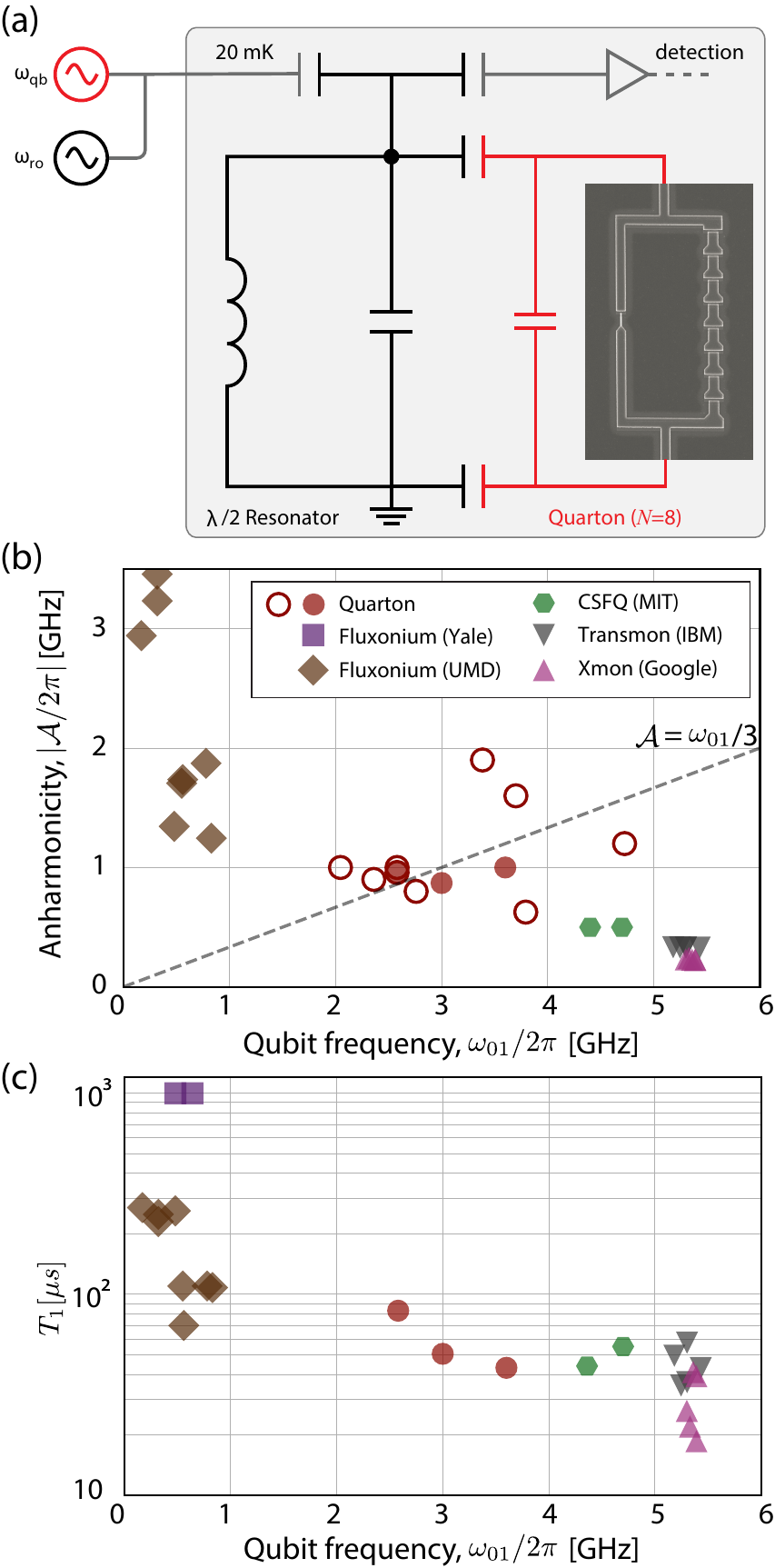}
		\caption[]
		{
			\textbf{(a)} Sketch of the measurement setup and SEM image of an 8-array-junction quarton qubit.
			\textbf{(b)} Comparison of anharmonicity and qubit frequency among different qubit variants.
			Results of all the quarton samples in Table~\ref{table1} are included.
			Solid filled circles are the devices shown in \textbf{(c)}.
			Fluxonium presented in Ref.~\cite{Pop-Nature-2014} have qubit frequencies about 300\,MHz but no information about anharmonicity.
			\textbf{(c)} Comparison of $T_1$ and qubit frequency.
			Only results from quarton samples with statistical confidence (Device A,C,I) are included.
		}
		\label{fig:modality_compare}
	\end{center}
\end{figure}

To demonstrate the concept, we implemented the quarton design with $N=8,16$.
As a practical matter, we have found it best to decide first on the target qubit frequency and its anharmonicity.
Since $\mathcal{A}/\omega_{01}$ mostly depends on $\gamma/N$, one may fix $\gamma/N$ before optimizing other parameters, simplifying the design process.
The circuit layout, fabrication process, and measurement setup are similar to those presented in our previous work \cite{Yan-NComms-2016}.
The aluminum metallization layer is patterned with square-shaped pads for the shunt capacitor and a half-wave-length transmission-line resonator for readout.
A slightly larger qubit loop, about $10\times20\,\mathrm{\mu m}^2$, is used here for housing the array junctions.
The junctions are made in the standard dolan-bridge style.

We tested multiple samples with varying parameters.
The typical design parameters are $I_\mathrm{C}=15-40$\,nA, $C_\Sigma=20-30$\,fF, $\gamma/N=0.85-1.1$.
Results are shown in Table I and Fig.~\ref{fig:modality_compare}.
Most samples have qubit frequencies
spread within 2-4\,GHz and anharmonicities above 800\,MHz, including some ideal cases like sample H ($\omega_{01}=3.4$\,GHz, $\mathcal{A}=1.9$\,GHz).
%
Fig.~\ref{fig:modality_compare}(b) shows that $\mathcal{A}/\omega_{01}$ ratios of these samples spread around the 1/3 line.
In comparison to transmon-type qubits and CSFQs which have much lower anharmonicities (200-300\,MHz and 500\,MHz respectively) and to fluxonium qubits whose qubit frequencies are consistently below 1\,GHz, the quarton qubits demonstrate a practically useful parameter regime where
trade-offs between qubit frequencies and anharmonicities can be made.

However, by comparing the predicted and experimentally inferred values of $\gamma/N$, we find that variation during fabrication and subsequent aging may cause significant fluctuations in actual values.
Qubit frequencies in many samples undershoot our target values ($\geqslant3$\,GHz), possibly due to junction-aging effects that reduce $I_\mathrm{c}$.
Studying and improving reproducibility will be a main objective in the future.

The quarton qubits show comparable $T_1$ times with respect to 2D transmon-type qubits and CSFQs [Fig.~\ref{fig:modality_compare}(c)].
We believe surface participation is the common key factor affecting coherence.
Fluxonium devices 
generally have longer $T_1$ times, in some cases on the order of 1\,ms \cite{Pop-Nature-2014}.
The enhancement is due to the suppressed dipole matrix element in the deep fluxon regime, and due to weaker low-frequency noise from Ohmic or super-Ohmic dielectric loss, i.e., the power spectral densities $S(\omega)\propto\omega^d$ ($d\geqslant1$) \cite{Nguyen_2019}.

The quarton qubits also show long spin-echo times $T_\mathrm{2Echo}$.
The highlighted samples in Table I approaches the $T_\mathrm{2Echo}\!=\!2T_1$ limit, indicating low residual thermal cavity photons due to our optimized measurement setup \cite{yan2018distinguishing}.
%
%

To conclude, our GFQ framework facilitates the understanding of how key qubit properties are related to circuit parameters.
In particular, we find the effectiveness of $\gamma/N$ in tuning the ratio between anharmonicity and qubit frequency and the effectiveness in suppressing charge dispersion by increasing $N$.
We experimentally demonstrate how to take advantage of these findings by testing the quarton design, which simultaneously achieves a desirable qubit frequency, large anharmonicity, and high coherence while maintaining a compact design.
The configurable energy level structure alleviates the problem of frequency crowding, promising better two-qubit gate performance from schemes such as 
parametric gates \cite{mckay2016universal,caldwell2018parametrically}.
Future improvement in reproducibility can transform such designs into powerful building blocks in quantum information processing.
%
%

\vspace{12pt}

\begin{table*}[t]
\begin{tabular}{ccccccccccc}
	\hline
	Device & $N$ & $I_\mathrm{c}$ [nA] & $C_\mathrm{sh}$ [fF] & $\gamma/N$ & $\omega_{01}$ [GHz] & $\mathcal{A}$ [GHz] & $\mathcal{A}/\omega_{01}$ & $T_1$ [$\mu$s] & $T_\mathrm{2Echo}$ [$\mu$s] & Local Bias \\
	\hline
	\rowcolor{green!25}
    A & 8 & 21 & 20 & 0.92 & 3.6 & 1.0 & 0.28 & $43.1\pm7.5$ & 70-100 & No \\
	\hline
	B & 8 & 21 & 30 & 0.95 & 2.8 & 0.8 & 0.29 & 23 & - & No \\
	\hline
	\rowcolor{green!25} C & 8 & 21 & 30 & 0.92 & 2.6 & 1.0 & 0.38 & $82.9\pm7.9$ & 100-125 & No \\
	\hline
	D & 8 & 18 & 30 & 0.92 & 2.4 & 0.9 & 0.38 & 20 & - & No \\
	\hline
	E & 8 & 18 & 30 & 0.93 & 2.0 & 1.0 & 0.50 & 50 & 12 & Yes \\
	\hline
	F & 8 & 40 & 20 & 0.93 & 3.7 & 1.6 & 0.43 & - & - & Yes \\
	\hline
	G & 8 & 40 & 20 & 0.98 & 4.7 & 1.2 & 0.26 & 10 & 7 & No \\
	\hline
	H & 8 & 40 & 20 & 1.0 & 3.4 & 1.9 & 0.56 & 23 & 15 & Yes \\
	\hline
	\rowcolor{green!25} I & 16 & 14 & 20 & 0.84 & 3.0 & 0.9 & 0.30 & $50.6\pm9.2$ & 110-140 & No \\
	\hline
	J & 16 & 14 & 20 & 1.09 & 3.8 & 0.6 & 0.16 & 30 & - & No \\
	\hline
	K & 16 & 27 & 20 & 0.88 & 2.6 & 1.0 & 0.38 & 30 & 20 & Yes \\
	\hline
	\rowcolor{blue!25} CSFQ & 2 & 60 & 50 & 1.2 & 4.7 & 0.5 & 0.11 & 35-55 & 70-90 & No \\
\hline
\end{tabular}
\vspace*{1mm}
\caption[]
{
	Design parameters ($N$, $I_\mathrm{c}$, $C_\mathrm{sh}$, $\gamma/N$) and measurement results ($T_1$ and spin-echo dephasing time $T_\mathrm{2Echo}$) of quarton qubits.
	The $\gamma/N$ ratios are designed to be in the vinicity of 1 with slight variations.
	We note that the measured frequencies mostly have better agreement with a 30-40\% lower $I_\mathrm{c}$ than the designed value, possibly due to junction aging.
	Highlighted are samples with repeated measurements of coherence times (typically 50-100 times over ~10 hours).
	Results of other samples lack of statistical confidence, and are presented for reference.
	The CSFQ results from Ref.~\cite{Yan-NComms-2016} are also listed for comparison.
	Note that all quartons have much smaller shunt capacitance than CSFQ.
	The reduced footprint promises better scalability.
}\label{table1}
\end{table*}

\section*{Acknowledgement}
We thank Andrew J. Kerman for helpful discussion. It is a pleasure to thank Mirabella Pulido for generous assistance. 
This research was funded in part by the U.S. Army Research Office Grant W911NF-18-1-0411; 
and by the Assistant Secretary of Defense for Research \& Engineering under Air Force Contract No. FA8721-05-C-0002. 
Opinions, interpretations, conclusions, and recommendations are those of the authors and are not necessarily endorsed by the United States Government.

\newpage
\bibliography{references}

\end{document}